\documentclass[journal=jacsat,manuscript=article]{achemso}
\usepackage[version=3]{mhchem} 
\usepackage{xcolor}

\author{Anand Dev Ranjan}
\email{adr20rs068@iiserkol.ac.in}
\affiliation[University]
{Department of Physics, IISER Kolkata, Mohanpur}
\author{Suyash Narayan Amzare}
\affiliation[University]
{Department of Health Technology, Technical University of Denmark (DTU), Denmark}
\author{Subhrokoli Ghosh}
\affiliation[University of Freiburg, Department of Microsystems Engineering - IMTEK]
{PicoQuant GmbH, Berlin, Germany}
\author{Ayan Banerjee}
\email{ayan@iiserkol.ac.in}
\affiliation[University]
{Department of Physics, IISER Kolkata, Mohanpur}

\title[An \textsf{achemso} demo]
  {Laser-Guided Microbubble Lithography for Multilayer Biophotonic Heterostructures}

\abbreviations{IR,NMR,UV}
\begin{document}

\begin{abstract}
The fabrication of multilayered heterostructures is pivotal in advancing microelectronics and biosensing platforms with enhanced sensing capabilities. Currently, these heterostructures are predominantly produced via top-down techniques, which are prohibitively expensive and impractical for biomedical applications. In this study, we demonstrate the fabrication of in-situ microstructures using the bottom-up approach we have developed, known as microbubble lithography. This technique involves the layer-by-layer self-assembly of various materials that adhere together, forming heterostructures. The resulting heterostructures were successfully employed for patterning a biomarker and a reporter protein, indicating the platform's suitability for sensing applications. This work paves the way for developing cost-effective and environmentally friendly sensing platforms using bottom-up fabrication methods.
\end{abstract}
\section{Introduction}
Heterostructures are three-dimensional assemblies where each layer is composed of different materials with unique properties and functions. They are crucial for developing microelectronic chips \cite{sun2021Wiley,ahmad2023Wiley,alferov2013IEEE} and biodetection platforms \cite{yu2024Wiley,wang2020RSC,anower2021Book,sakthivel2023Elsevier}, which require the precise stacking of materials to create functional layers. These structures are widely used in electronic and optoelectronic devices, including transistors \cite{iannaccone2018Nature}, lasers\cite{lyu2023Wiley}, light-emitting diodes (LEDs)\cite{chen2021Small}, photodetectors\cite{chen2020Wiley,cheng2019Wiley}, and biosensors\cite{liu2022Wiley,du2023Wiley}, due to their ability to combine materials with specialized capabilities. Fabrication methods for heterostructures include chemical vapor deposition (CVD), pulsed laser deposition (PLD), sputtering, and lithographic techniques like photolithography and electron beam lithography. Although methods like CVD and lithography are scalable, they often involve complex and costly additional processing steps. Self-assembly techniques, such as solvent annealing, solvent evaporation, and controlled aggregation, offer solutions to these challenges by simplifying the manufacturing process with minimal processing. Out of them, directed controlled aggregation has the potential to produce complicated structures with microscale accuracy. However, hurdles persist in achieving a functionalized heterostructure for direct sensing applications. There have been some accomplishments in this respect, such as hierarchical self-assembly approach by tuning the noncovalent interactions for the longitudinally-epitaxial growth (LG)\cite{zhuo2019Nature}, 3D nanoprinting technology, which uses two-step absorption via an integrated fibre linked diode to manufacture sub-micron structures\cite{liu2023OSA}, electrostatic self-assembly\cite{lobo2024RSC} and so on. Some of these methods do harness the capability of the bottom-up approach but they also have additional post-processing steps. Additionally, these often lack the ability to form well defined structures as they are self-assembly based, with limited external control.

Microbubble lithography (MBL) is a decade old self-assembly technique which uses a laser-induced microbubble for fabricating microstructures \cite{ghosh2023nl}. It has been used for multifarious applications ranging from designing complex architectures with quantum dots \cite{}, electronic chips using metal nanoparticles \cite{} and conducting polymers \cite{}, site-specific micro-catalysis \cite{}, and recently, biosensing \cite{}. Especially in the latter, MBL -- which is a self assembly-based approach -- appears promising in the development of heterostructures since the self-assembly it drives is essentially directed, and thus capable of producing defined structures in real time. However, there remain certain challenges in creating a heterostructure employing MBL. These can be described in the following manner: 1) Nucleation of the microbubble: the very first prerequisite for microbubble-based self-assembly is the formation of microbubble which requires the presence of absorbing particles/substrates, and since heterostructures are comprised of diverse materials that may not be absorbing at the specified wavelength and hence may not form a microbubble. 2) Adherence of layers: multiple layers may not adhere to one other. 3) Temperature: this is one of the biggest concerns while patterning any biologically active molecule as high temperature is detrimental for both proteins and biopolymers. Recently several groups have tried to solve these challenges by generating a microbubble at low temperature \cite{}, or using shrinking microbubbles \cite{}. Now, while the production of conductive patterns using MBL has been pursued for sensors using multilayered structures \cite{}, the development of some of the layers rely on other methods, including the deployment of unusual solvents or additional post-processing procedures, limiting the adaptability of MBL. Indeed, to the best of our knowledge, there is no evidence of the fabrication of heterostructures entirely deploying MBL. It is clear that extending the capabilities of MBL towards the devleopment of heterostructures would establish this technology as a robust enabler for biosensing platforms and microelectronics devices, requiring minimal postprocessing.

This work focuses on the creation of a multilayer platform composed of  three different layers of materials employing MBL. The first layer is chosen for its high absorptivity at the laser wavelength, allowing it to serve as a scaffold for the microbubble formation. This allows us to choose the second layer independent of its absorbance at the given wavelength. Consequently, anything in the dispersed phase can self-assemble on top of the fabricated first layer. This eliminates the problem of microbubble nucleation even for non-absorbing materials at the specified wavelength, as well as improving the site-specific assembly of the subsequent layers, which only get assembled at the scaffold's position due to the formation of microbubbles only at those places. Although patterning scaffolds address the issue of microbubble formation and directed self-assembly, they do not guarantee that the self-assembled second layer will attach to the targeted positions. Hence, we use cross-linking agents in the secondary layer to promote layer adhesion. As a result, the cross-linker should be chosen in accordance with the final layer. This layer serves as a bridge connecting the primary layer to the final layer of materials, resulting in a sandwich type configuration. Finally, depending on the cross-linker used, anything can be chosen as the tertiary layer, which automatically combines with the secondary layer resulting in a 3 layer stacked heterostructure. This tertiary layer can be any analyte that changes its response under the exposure of the material to be detected. In this work, we create Ammonium tetrathio molybdate (ATTM) patterns to serve as the first layer for a patterning scaffold. Because our ultimate goal is to pattern biomolecules, we used APTES, a well-studied cross-linker, as our second layer. Finally, we pattern a variety of biomolecules on top of these layers to create a heterostructure platform. The multiple experimental parameters required for rapid and consistent self-assembly of colloidal particles on a substrate are determined. To identify the optimal conditions for heterostructure development, colloidal concentrations were tested for all of the layers with varying laser intensity. The resulting heterostructures were shown to be highly selective and capable of detecting model analytes at micromolar concentrations. Overall, this study sets the groundwork for the manufacture of multilayer heterostructures using microbubble lithography, which should enhance the manufacturing of MBL-based sensors and microelectronics and promote applications requiring highly directed and selective self assembly.



\section{Results and Discussions}
\begin{figure}[!h]
\centering
\includegraphics[width=16cm,height=5cm]{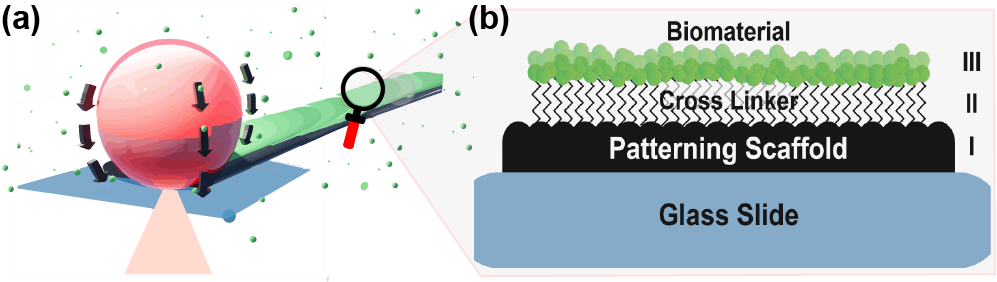}
\caption{(a) Self assembly of materials employing microbubble lithography (MBL) where the different materials are self-assembled on top of each other (b) Different layer of materials self-assembled using MBL.}
\label{Heterostructure_setup}
\end{figure}

We fabricate the heterostructures using an Olympus 1X71 microscope with an overfilled objective. The 100x objective focuses the 1064 nm laser on the sample plane resulting in a spot size of 3 um. The sample chamber is filled with nearly  10$\mu$l volume of the ATTM dispersion. As the dispersion has very high absorptivity at the laser wavelength as shown by the \ref{FigS1(B)}, hence the light absorbed by the ATTM microparticles results in the formation of a microbubble, which coupled with the stage movement, results in the formation of various patterns as shown in the schematic Fig.~\ref{Heterostructure_setup}(a). This can be used to make layer by layer assembly of various materials on top of each other as shown in Fig.~\ref{Heterostructure_setup}(b). The typical experimental arrangement used in the setup is similar to the used in our previous work [Small 2024]. 

Now, in order to form a heterostructure -- the biggest problem that arises is the absorptivity of the material to be patterned. Usually, in MBL -- as reported in the literature -- only those materials which are highly absorptive at the laser wavelength. are patterned However, this poses a serious limitation in the flexibility and cost-effectiveness of the technology, since it  necessitates the use of lasers of different wavelengths to pattern different types of materials. Indeed, such issues would be even more relevant for materials having absorptivity in the UV range. The question thus arises -- if the material to be patterned is not absorptive at the laser wavelength, then is it not possible to form continuous self assembly of the material? One of the possible solutions is the use of absorptive substrates, but even that adds a certain complexity and cost associated with the nontrivial surface engineering required to induce the required plasmonic excitation (ref substrate based work). Here, we report a unique method of immobilising any material on top of pattern formed by the MBL, which is achieved by using a cross-linker which attaches the patterned scaffold to the any analyte which is needed to be immobilised.

\subsection{First layer: Patterning Scaffold}
In order to construct a heterostructure, the base pattern must be consistent and continuous since it serves as a scaffold for the attachment of the other layers. As a result, optimising this layer is critical to the creation of the heterostructure. The laser power, particle concentration, and patterning speed are the three criteria that determine pattern quality in MBL. Thus, we pattern ATTM on glass slides, and statistically modify these parameters to get the ideal combination for producing uniform and continuous patterns.

\subsubsection{Intensity}
\begin{figure}[!h]
\centering
\includegraphics[width=16cm,height=11cm]{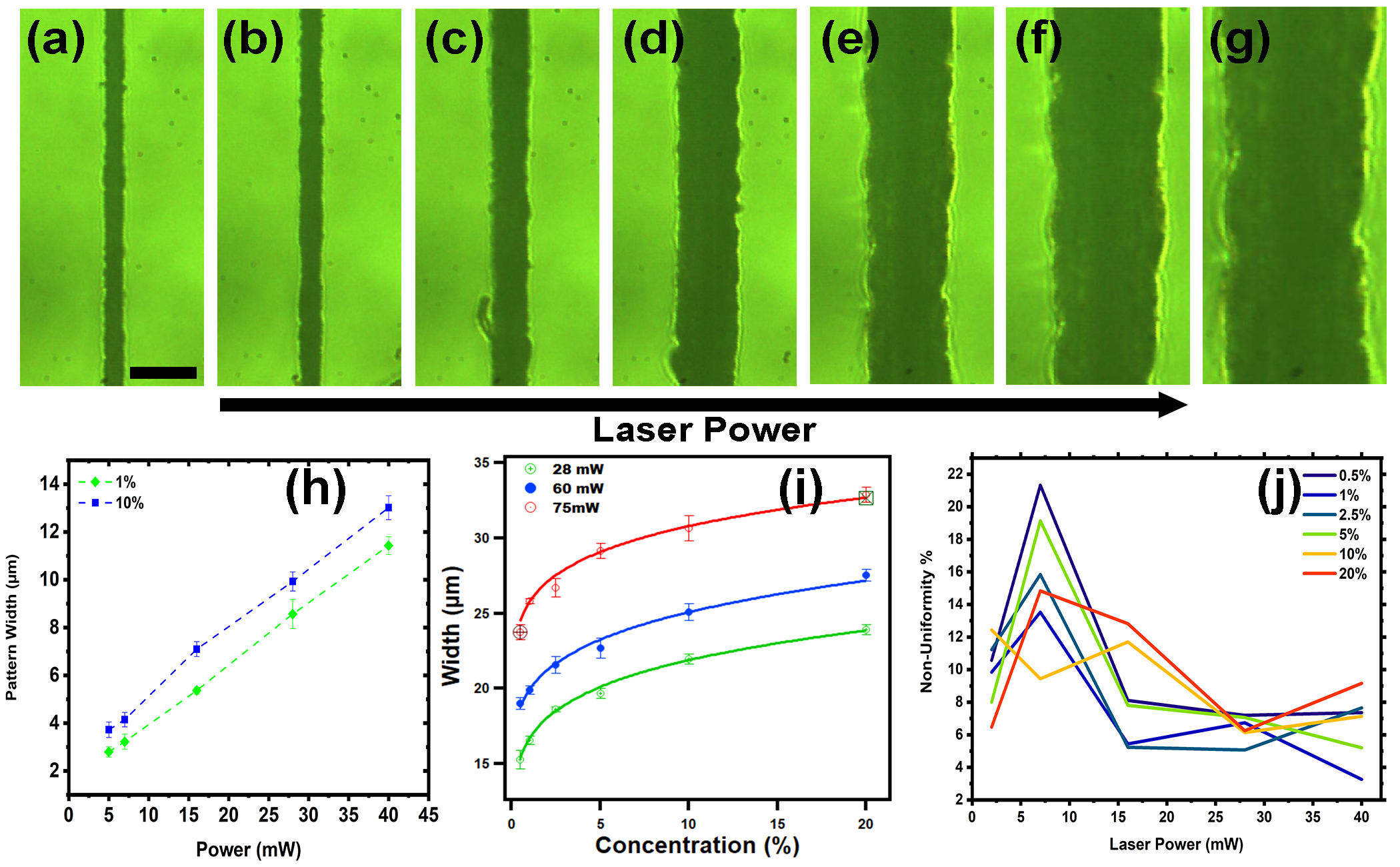}
\caption{(a-g) Increase in the width of the fabricated pattern with increasing laser power at 5$\%$ concentration of ATTM (scale bar represents 10$\mu$m (h) The linear increase in the width of fabricated patterns with increasing laser power (i) Width of the patterns at different concentration shows a power law behaviour and the graph saturates at higher concentration values for each laser power (j) Non-uniformity of the patterns formed at different laser powers and concentrations.}
\label{ATTM-Optimisation}
\end{figure}

We pattern ATTM at different laser powers starting from 2 mW to 40 mW present at various concentrations (Fig.~\ref{ATTM-Optimisation}(a-g)). As clear from the Fig.~\ref{ATTM-Optimisation}(h), irrespective of the concentration of the dispersion, the width of the patterns  increases with increase in laser power. This is due to the fact that a rise in laser power increases the size of the microbubble in accordance with Fourier's Law\cite{liu2015Nature}: 
\[ \frac{dE}{dt} = 4\pi kR^2\delta_r T \propto R \]
The problem's spherical symmetry leads to temperature being spherically symmetric, which means that the radius of the microbubble should increase linearly with the generated heat or laser intensity. This in turn results in enhanced self assembly that causes higher deposition of the material around the bubble base. 

\subsubsection{Concentration}
We also investigated the role of concentration on the uniformity of patterns. Interestingly we found that the pattern width versus concentration follows a power law with an average range of exponents between 0.15 to 0.2, with an error of around 0.xx, as shown in Fig.~\ref{ATTM-Optimisation}(i). As is clear from all the plots at different powers, the width of the patterns starts saturating at a given laser power as the concentration increases. This seems intuitive as for a given power, increasing the concentration starts increasing the volume of particles getting self-assembled. If we further increase the concentration, since the maximum bubble size has already been attained, and only gaps in the self-assembled particles can be filled. Thus, with increasing concentration - the width does not increase. However, the height of the patterns are observed to increase linearly (WE NEED TO CHECK THIS OUT!)...

\subsubsection{Uniformity}
We further checked the uniformity of these patterns formed at various laser powers and concentrations. The most important factor that affects the uniformity of pattern widths is the constancy of bubble size as it is translated - any variations would lead to corresponding changes in pattern width. Out of the various concentrations we tested -- for laser powers below 5 mW --  all concentrations did not lead to uniform patterns (average variation in width more than xx \%). This is expected since this is close to the minimum power required for a microbubble to form, so that as the bubble was translated there could be variations in size with the laser power not being able to sustain a particular size of the bubble. However, at laser powers higher than 15 mW, the 1\%, 2.5\%, and 5\% solutions showed the most uniformly wide patterns, with variations less than xx\%. The 10\% solution also showed good results, but the issue in this case that the dispersion was so densely populated with micro-particles that visibility and washing post-patterning become major issues. The micro-bubble not being clearly visible while patterning results in uncertainty over the speed required to move the microscope translation stage, which is a critical factor in forming the patterns, as well as ensuring their uniformity. Therefore, a concentration in the range of 1\% to 5\% seems to be the best for pattern uniformity and ease of patterning.

\subsection{Second Layer: Cross-Linking Agent}
As mentioned earlier, the first layer (also called base layer) was created using 1\% ATTM at 30mW laser power, and served as the basic layer for constructing any type of heterostructure. The use of ATTM as the base layer eliminated the need to use different lasers of wavelengths chosen according to the absorptivity of the second material to be patterned, since microbubbles can form on top of this layer at rather low laser powers (around 5 mW) without affecting its structural integrity. Indeed, the formation of microbubbles on top of this layer leads to assembly of everything dispersed in the fluid surrounding the layer. We employed APTES, a crosslinking agent that can be used to interconnect the base and third analyte layers. Since the APTES layer was colourless, standard brightfield microscopy, as employed for the ATTM layer, was inconclusive for its characterisation. As a result, this APTES layer was characterised using Atomic Force Microscopy (AFM) and X-ray Photoelectron Microscopy (XPS). AFM describes pattern homogeneity, while XPS validates the presence of amine moeities on the glass slide's surface.

AFM images of the bare ATTM pattern and APTES-coated ATTM (ATTM-APTES) patterns are shown in Fig 3(a-b). The AFM images distinctly show that the ATTM-APTES patterns contain granular particles on the surface, whereas the bare ATTM pattern exhibits the characteristic hump geometry as shown in Fig 3(c), consistent with previous observations of patterns formed using MBL (ref). We hypothesize that the granular structures are due to the adhesion of APTES on the ATTM surface, which causes the surface to become rougher with pronounced crests and troughs. These features provide an ideal environment for proteins or other biomolecules to concentrate locally, potentially leading to increased capture of biomolecules on these specifically designed patterns.   

\begin{figure}[!h]
\centering
\includegraphics[width=16cm,height=5cm]{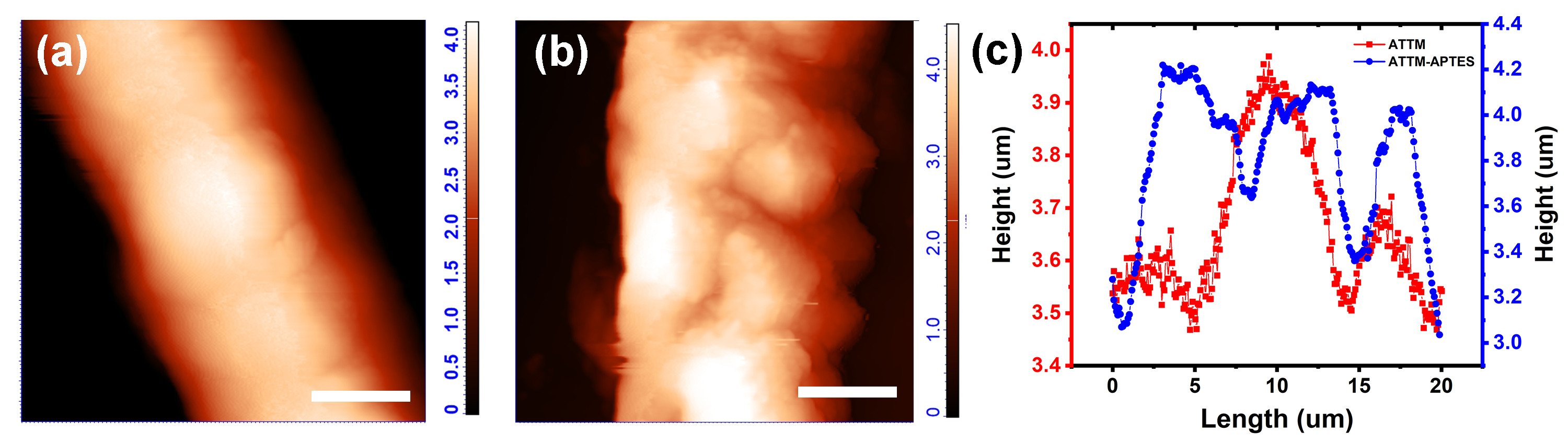}
\caption{(a) AFM images of the ATTM patterning scaffold shown uniformly fabricated patterns while (b) AFM images of the ATTM-APTES layer shown non-uniform surface with granular boundaries (c) ATTM patterns shows the typical hump geometry whereas the ATTM-APTES shows the multiple crests and troughs ideal for capturing sub-micron particles.}
\label{AFM}
\end{figure}

\begin{figure}[!h]
\centering
\includegraphics[width=16
cm,height=8cm]{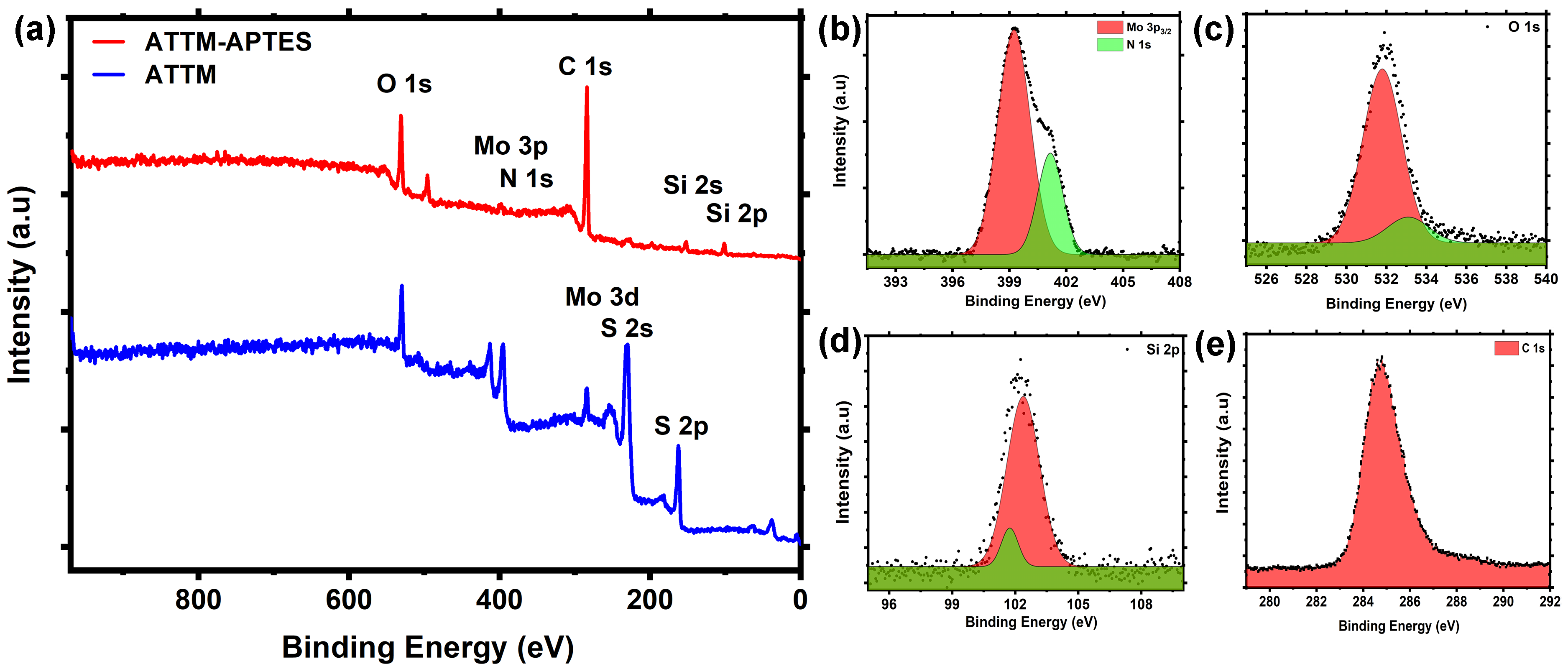}
\caption{(a) The XPS survey spectra of ATTM pattern (b) High resolution XPS spectra of the N, O, Si and C present in the ATTM-APTES pattern.}
\label{XPS}
\end{figure}

XPS was used to examine the surface electronic states and chemical compositions of ATTM and APTES. The survey spectra of the ATTM (Fig.~\ref{XPS}(a)) establish the presence of Mo, N, S, O, and extra peaks of C and Si in the case of ATTM-APTES. The narrower cross-section of N1s compared to Mo3p and other core levels resulted in no readily apparent signal for N in the ATTM-APTES survey. Nonetheless, the high resolution spectra revealed a strong peak at 401.1 eV, close to the Mo3p level, for N1s in the amine group (Fig 3b). The presence of these amine moeities enhances the surface capture of biomolecules and therefore provide an ideal ground for the their immobilisation. The deconvouluted high-resolution O1s XPS spectra of ATTM-APTES samples reveal distinct oxygen peaks at 531.8 eV and 533.1 eV. These peaks are indicative of the oxygen atoms in SiO$_2$ and the surface hydroxyl or carboxyl groups (OH/CO) present on the ATTM-APTES surface. Furthermore, the SiO$_2$ peak's presence in the Si2p spectra, observed at 102.4 eV and 101.7 eV, aligns with the hypothesized formation of silane cross-linking molecules on the ATTM surface.
 
The XPS and AFM clearly establishes the attachment of the cross-linker molecule on top of the patterning scaffold. Finally to establish this configuration we need to check the attachment of a reporter biomolecule on top of these patterns. Hence, the final layer was made of the model analyte which shows some form of visible change if reacted by the sensing agent. We chose to work with Rhodamine since it is a fluorophore that has wide applications ranging from material sciences to biology. 

\subsection{Third Layer: Biomarker}
\begin{figure}[!h]
\centering
\includegraphics[width=16cm,height=7cm]{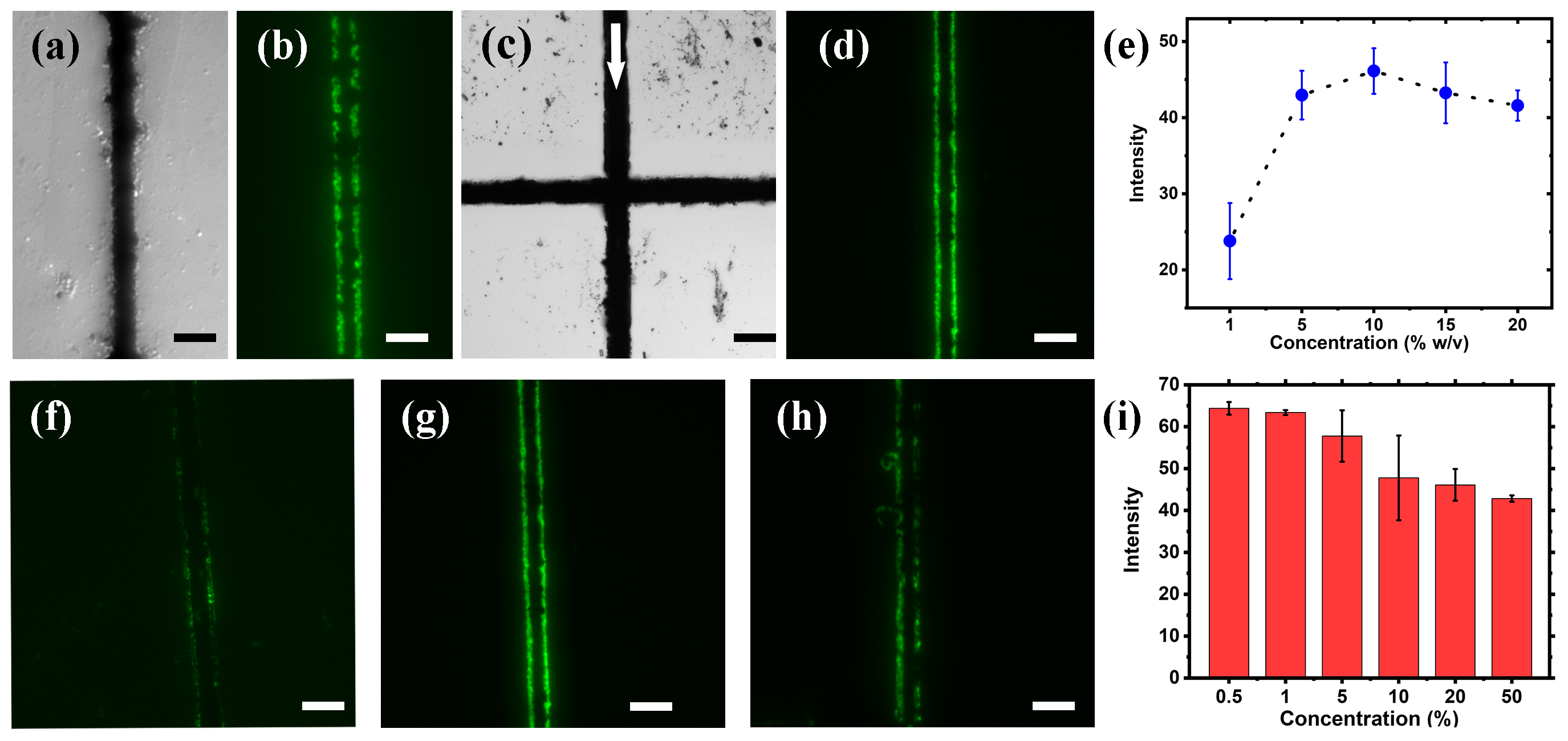}
\caption{(a-b) DIC and corresponding fluorescence images of the 3 layer heterostructure (c-d) DIC and corresponding image of the ATTM patterned in a cross geometry where microbubble is redrawn only in the vertical direction (as shown by arrow) for the second and third layers (e) change in the fluorescence intensity of the rhodamine when patterned at various concentrations (f-h) patterning of 10$\%$ rhodamine at 0.25$\%$, 1$\%$ and 50$\%$ concentrations of APTES respectively (i) change in the fluorescence intensity of the final layer of rhodamine when patterned at various concentrations of APTES.}
\label{Third-Layer}
\end{figure}

Here, we repeated the entire patterning procedure on top of the previously described patterns using two layers of material. Thus, our patterning scaffold (ATTM) was the first layer, followed by the cross-linker (APTES), while Rhodamine was the third layer. Figure ~\ref{Third-Layer}(a-b) displays brightfield and fluorescence pictures of immobilised rhodamine. Clearly, the fluorescence spectra reveal that rhodamine is immobilised only on the patterned scaffold area, demonstrating excellent selectivity of our biomolecule capture approach. The question however arises whether MBL was essential for the immobilisation of biomolecules at our length and time scales.  In order to validate this, we constructed the ATTM patterns in cross-geometry as shown in Fig.~\ref{Third-Layer}(c). For the second and third layers, microbubbles were translated solely in the vertical direction, as demonstrated by the arrow in Fig.~\ref{Third-Layer}(c). The fluorescent image in Fig.~\ref{Third-Layer}(d) indicates the robustness of the MBL process, as Rhodamine only adheres to MBL patterning locations. As a result, we demonstrated that at the timescales of our experiment, the MBL process is important for immobilising a given biomolecule. To confirm this fact further, we tested whether the attachment worked with different concentrations of the final layer. Fig.~\ref{Third-Layer}(e) shows that the attachment of Rhodamine decreases at lower concentrations and saturates as the concentration increases. This may be due to a decrease in the number of accessible binding molecules at lower concentrations. As the concentration increased, the majority of the binding sites were filled, resulting in saturation of the fluorescence signal. 

We also conducted an optimisation study of the concentration of APTES on the final layer for the most robust biomolecule adhesion. This was accomplished by measuring the adherence of the final layer (rhodamine analyte) to various APTES concentrations. Towards this, we created the cross-linker layer with varying APTES concentrations (0.5$\%$-50$\%$) at a fixed Rhodamine concentration (10$\%$). We discovered that higher doses of APTES enhanced the overall intensity of the fluorescence signal, which includes both signal and background (error bar increases). This is because high APTES concentrations allow for non-uniform biomolecule attachment at many accessible sites, limiting control over immobilisation at specific sites. As a result, it boosts both the background and the fluorescence signal, resulting in a low signal-to-noise ratio. Thus, 0.5 to 1 $\%$ showed optimal normalised fluorescence intensity for the patterned Rhodamine. 

While the experimental fluorescence patterns confirm robust and selective immobilisation of Rhodamine on APTES-modified ATTM scaffolds, the underlying spatial distribution of biomolecules—particularly the observed ring-like intensity profile—warrants further understanding. To investigate the physical mechanisms driving this distribution, we developed a phenomenological model based on radial protein concentration gradients induced by the microbubble. The model captures the thermophoretic depletion near the bubble center and the peripheral enrichment at the triple-phase contact line, where APTES promotes immobilisation.

\subsubsection{Simulated Protein Distribution: Double-Gaussian Model}
To understand how proteins like Rhodamine 6G spatially accumulate on the laser-induced pattern, we developed a 2D radial simulation based on the expected thermophoretic and binding interactions near the microbubble. The temperature profile around a laser-induced microbubble is radially symmetric, with a high peak at the center and a gradual decay toward the periphery. Proteins tend to denature or desorb at elevated central temperatures. The periphery of the bubble is relatively cooler, favoring stable binding. The environment is cylindrically symmetric. That is, there's no preferential direction—what matters is the radial distance r from the center. So, any realistic model must be radial and symmetric around the center:
\begin{equation}
    B(r)=f(r)    
\end{equation}
A Gaussian function naturally models radial decay or rise in properties like concentration or temperature. The first Gaussian models suppression of protein binding due to high temperatures or material degradation at the bubble center:
\begin{equation}
    C_1(r) = A_1 e^{\frac{r^2}{2\sigma_1^2}} 
\end{equation}
with $A_1 < 0$, this term peaks negatively at the center, creating a well in binding occupancy. The second Gaussian accounts for enhanced adsorption near the bubble boundary, which occurs because: Temperature is optimal for protein stability, APTES surface is intact and Convection and thermophoresis concentrate proteins at the edge. So we place a Gaussian centered at the bubble radius R, e.g.,
\begin{equation}
    C_2(r) = A_2 e^{\frac{(r-R)^2}{2\sigma_2^2}}
\end{equation}
This results in a binding suppression at the center and enhanced adsorption near the bubble edge. Given that the microbubble center reaches the highest temperature, proteins tend to deplete at the center and concentrate more at the cooler periphery, particularly where APTES is present. We modeled the protein concentration profile as a double-Gaussian distribution, described by:
\begin{equation}
    C(r) = C_1(r) + C_2(r) = -|A_1| e^{\frac{r^2}{2\sigma_1^2}} + |A_2| e^{\frac{(r-R)^2}{2\sigma_2^2}}
\end{equation}
: parameters for the inner depletion zone (center of the bubbe: parameters for the outer accumulation zone near the triple-phase contact line (TPCL).

The simulation shows a pronounced minimum at the bubble center and peak intensity at the outer rim, matching both thermophoretic expectations and experimental fluorescence profiles. This explains the ring-like fluorescence patterns often observed in high-resolution images of Rhodamine-labeled layers.

\begin{figure}[!h]
\centering
\includegraphics[width=16cm,height=4cm]{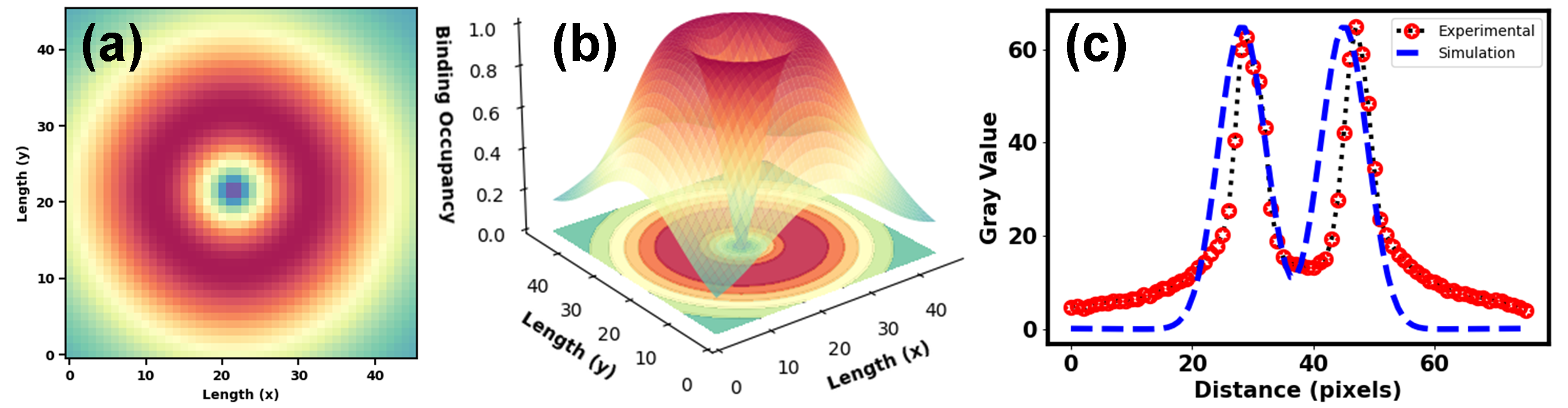}
\caption{(a) Increase in the width of the fabricated pattern with increasing laser power at 0.5$\%$ concentration of ATTM (scale bar represents 10$\mu$m (b) attachment of fluorophore at 50\% concentration of APTES (c) ATTACHMENT OF rhodamine at 0.5\% of the APTES solution }
\label{APTES-RH-Optimisati}
\end{figure}

The results further validate our three-layered sandwich model by showing that the protein immobilization is not uniform but radially modulated, with optimal capture at the APTES-rich periphery. It is also evident, that MBL patterns display a gradient in terms of analyte binding with higher binding near the edges, which would be an important factor to consider in designing biosensors using this technology.

Finally, we fabricated a 3 layered heterostructure where APTES (layer 2) acted as as a cross-linker between the patterning scaffold (layer 1, ATTM patterns) and the a model analyte (layer 3), Rhodamine 6g. This results in a three layer heterostructure where each layers were stacked on top of each other resulting in a sandwich type configuration as shown in Fig 1(b). But why we have no fluorescence at the center of the pattern ? 



\subsection{Conclusion}
In this work, we attempt to understand the novel and entirely optically driven method for the construction of multilayered biofunctional heterostructures using MBL, tailored for next-generation biophotonic and optoelectronic interfaces. Our three-tiered fabrication approach—combining a plasmonic ATTM scaffold, an APTES cross-linker, and a protein or fluorophore top layer -- mitigates key limitations of conventional patterning techniques, including harsh processing conditions, limited material compatibility, and spatial inflexibility. By leveraging the precise flow dynamics induced by plasmonic microbubbles and tuning experimental parameters such as laser power, particle concentration, and cross-linker density, we achieved robust, spatially defined architectures of the well-known fluorophore Rhodamine 6G with sub-10 µm resolution. Notably, our results show that reduced-power microbubble translation preserves the fluorophore functionality while enabling selective immobilization, as verified through fluorescence imaging. The emergence of a distinctive double-hump fluorescence intensity profile —- well-described by a double-Lorentzian model -- underscores the nuanced interplay between thermal gradients, Marangoni flows, and surface-mediated anchoring at the triple-phase contact line. This unique distribution of the patterned fluorophore reinforces the utility of MBL as a platform not just for structural assembly, but also for fine control of molecular localization. 
Overall, this work confirms MBL as a scalable, mask-free, and biofriendly route for constructing chemically stable and spatially programmable biointerfaces. It opens a versatile avenue for integrating biological functionality into optical and electronic microsystems -- laying the groundwork for advanced biosensors, lab-on-chip diagnostics, and bio-optoelectronic hybrid devices in the emerging landscape of biophotonics.






\section{Experimental}
We use Olympus IX71, coupled with a CW IR laser (1064 nm) for the formation of the heterostructure. IR laser is tightly focused at the interface of the glass and the colloidal dispersion of ATTM placed on a glass slide. A vapour micro-bubble is nucleated on the surface, invoking the Marangoni flow around it. ATTM particles are assembled at the bottom of the bubble and are deposited there. The laser spot is moved around by translating the slide mount using a joystick control. This forms a continuous pattern of deposited
ATTM on the surface.

(3-Aminopropyl)triethoxysilane is a silane that provides covalent attachment of organic films to metal oxides. It will be used as an intermediate material here to attach
the Rh6 with patterned ATTM. APTES has alkoxy groups that get hydrolyzed in
water. A solution of APTES in ethanol is used in this step. The same process is
repeated here, however with a low laser power ($\sim$ 2mW). The micro-bubble is nucleated again and directs the self-assembly of APTES onto the patterned ATTM.
The volatile nature of ethanol aids in washing away the rest of the colloid after
re-patterning.

In this project, the same procedure is used to attach Rhodamine 6G (the material of
interest here) in an aqueous solution, with an intermediate laser power ($\sim$ 3.8mW). On self-assembly, the dye covalently bonds to the immobilized APTES on the pattern
after the re-patterning it, too, on the same part of the pattern where the APTES
was self-assembled.
Thus, we have successfully turned the directed self-assembly of the Rhodamine
6G on glass from reversible to irreversible. These heterostructures are washed with
distilled water and phosphate buffer solution (PBS). However, there are variables
that affect the overall quality of the heterostructures. These need to be optimized
to achieve high robustness and long life"

\begin{acknowledgement}
The authors thank IISER Kolkata, an autonomous research and teaching institution funded by the MHRD, Government of India for providing the financial support and infrastructure. 
\end{acknowledgement}


\bibliography{acs-achemso}

\end{document}